\documentclass[aps,prd,reprint,amsfonts,amssymb,amsmath showpacs,
preprintnumbers,letterpaper,superscriptaddress,nofootinbib,urlcolor=black,
linkcolor=black,twocolumn]{revtex4}

\usepackage{graphicx,epsfig,epsf}
\usepackage{slashed}
\usepackage{mathrsfs,mathtools}
\usepackage{dcolumn}
\usepackage{appendix}
\usepackage{verbatim}
\usepackage{psfrag}
\usepackage{bm}
\usepackage[utf8]{inputenc}
\usepackage{soul}
\usepackage{hyperref}

\def\half{{\textstyle{\frac{1}{2}}}}

\def\cPT{$\mathcal {PT}$}

\def\e{\mathrm{e}}
\def\d{\mathrm{d}}
\def\i{\mathrm{i}}
\def\P{\mathcal{P}}
\def\D{\mathcal{D}}
\def\T{\mathcal{T}}
\def\C{\mathcal{C}}
\def\J{\mathcal{J}}

\def\half{\textstyle{\frac{1}{2}}}
\def\fourth{\textstyle{\frac{1}{4}}}

\begin{document}

\preprint{KCL-PH-TH-2022-25}

\title{\cPT-symmetric $-g \varphi^4$ theory}

\author{Wen-Yuan Ai}
\email{wenyuan.ai@kcl.ac.uk}
\affiliation{Theoretical Particle Physics and Cosmology Group, Department of Physics, King's~College~London, Strand, London WC2R 2LS, UK}

\author{Carl M. Bender}
\email{cmb@wustl.edu}
\affiliation{Department of Physics, Washington University, St. Louis,
Missouri 63130, USA}

\author{Sarben Sarkar}\email{sarben.sarkar@kcl.ac.uk}
\affiliation{Theoretical Particle Physics and Cosmology Group, Department of Physics, King's~College~London, Strand, London WC2R 2LS, UK}


\begin{abstract}
The scalar field theory with potential $V(\varphi)=\textstyle{\frac{1}{2}} m^2\varphi^2-\textstyle{\frac{1}{4}} g\varphi^4$ ($g>0$) is ill defined as a Hermitian theory but in a non-Hermitian $\mathcal{P}\mathcal{T}$-symmetric framework it is well defined, and it has a positive real energy spectrum for the case of spacetime dimension $D=1$. While the methods used in the literature do not easily generalize to quantum field theory, in this paper the path-integral representation of a $\mathcal{P}\mathcal{T}$-symmetric $-g\varphi^4$ theory is shown to provide a unified formulation for general $D$. A new conjectural relation between the Euclidean partition functions $Z^{\mathcal{P}\mathcal{T}}(g)$ of the non-Hermitian $\mathcal{P}\mathcal{T}$-symmetric theory and $Z_{\rm Herm}(\lambda)$ of the $\lambda \varphi^4$ ($\lambda>0$) Hermitian theory is proposed: $\log Z^{\mathcal{P}\mathcal{T}}(g)=\textstyle{\frac{1}{2}} \log Z_{\rm Herm}(-g+{\rm i} 0^+)+\textstyle{\frac{1}{2}}\log Z_{\rm Herm}(-g-{\rm i} 0^+)$. This relation ensures a real energy spectrum for the non-Hermitian $\mathcal{P}\mathcal{T}$-symmetric $-g\varphi^4$ field theory. A closely related relation is rigorously valid in $D=0$. For $D=1$, using a semiclassical evaluation of $Z^{\mathcal{P}\mathcal{T}}(g)$, this relation is verified by comparing the imaginary parts of the ground-state energy $E_0^{\mathcal{P}\mathcal{T}}(g)$ (before cancellation) and $E_{0,\rm Herm}(-g\pm {\rm i} 0^+)$. 
\end{abstract}

\maketitle

\section{Introduction}

In search of physics beyond the Standard Model of particle physics, non-Hermitian \cPT-symmetric Hamiltonians have been used recently in model
building \cite{R01}. The importance of \cPT~symmetry for non-Hermitian
quantum-mechanical theories with real energy eigenvalues was discovered in
1998~\cite{R02} on the basis of numerical and semiclassical arguments. The
quantum-mechanical theories considered are governed by a Schr\"odinger equation
with potential $V(x)=m^2x^2/2+gx^2(\i x)^\epsilon/4$, where $\epsilon$ is a real
parameter. This potential is invariant under \cPT~reflection $\P:\ x\to-x,\
\T:\ x\to x,\ \i\to -\i$. For $\epsilon\ge0$ the energy spectrum of the
Hamiltonian was found to be real numerically~\cite{R02}. For the massless case
spectral reality was proved for $\epsilon>0$ by Dorey {\it et al.} using the
methods of integrable systems~\cite{R03,R04}. For $\epsilon=2$ the massless case can be mapped to a Hermitian Hamiltonian with the same spectrum~\cite{R05,R06}. The study of \cPT-symmetric
systems is an active research area~\cite{R04,R07}.

Quantum mechanics is quantum field theory in one-dimensional spacetime. However,
very little is known about the nature of \cPT-symmetric quantum field theory
with spacetime dimension $D>1$, and the calculational tools required for
analyzing such theories remain largely undeveloped. Methods that have been
successful in $D=1$~\cite{R04} do not extend to higher dimensions. Recent
papers~\cite{R08,R09,R010} have considered the case $D>1$, but the
proposed methods have yet to be implemented in $D=4$.

In this paper we study the non-Hermitian \cPT-symmetric scalar theory with potential $V(\varphi)=m^2\varphi^2/2-g\varphi^4/4$ ($m,\,g>0$) using the
path-integral formulation of quantum theories. Unlike earlier methods, here we
study this theory by constructing a relation between its Euclidean partition
function and that of a Hermitian theory with a positive quartic potential
$V(\phi)=m^2 \varphi^2/2+\lambda\varphi^4/4$ ($m,\,\lambda>0$). Specifically, we
conjecture and propose the following general relation for $D\geq1$:
\begin{eqnarray} 
\log Z^{\P\T}(g)&=&\half\log Z_{\rm Herm}(\lambda=-g+\i 0^+)\nonumber\\
&&\quad+\half\log Z_{\rm Herm}(\lambda=-g-\i 0^+),
\label{e1}
\end{eqnarray}
where $Z_{\rm Herm}(\lambda)$ is the usual Euclidean partition function for the
Hermitian theory, which must be analytically continued to $\lambda=-g\pm\i 0^+$
($g>0$) in this equation. The functional integral $Z^{\P\T}(g)$ must be defined
with an appropriate contour $\C_{\P\T}$ to ensure both existence and \cPT~symmetry, as will be explained. The Euclidean partition function can always be
put in a form of an exponential function via the exponentiation of disconnected
Feynman diagrams in perturbation theory or of the multi-instanton contributions
when there are nonpeturbative stationary points; $\log Z$ is related to the
energy. Therefore, \eqref{e1} implies that
\begin{align}
\label{e2}
\!E^{\P\T}(g)=\half E_{\rm Herm}(-g+\i 0^+)+\half E_{\rm Herm}(-g-\i 0^+).
\end{align}
For positive real $\lambda$, $E_{\rm Herm}(\lambda)=E^*_{\rm Herm}(\lambda^*)$.
This relation is analytic, so it holds when $E_{\rm Herm}(\lambda)$ is
analytically continued to $\lambda=-g+\i 0^+$. We thus have ${\rm Re}[E_{\rm
Herm}(-g+\i 0^+)]={\rm Re} [E_{\rm Herm}(-g-\i 0^+)]$ and ${\rm Im}[E_{\rm Herm}
(-g+\i 0^+)]=-{\rm Im} [E_{\rm Herm}(-g-\i 0^+)]$. Therefore, \eqref{e1}
implies a real energy spectrum for the non-Hermitian $-g\varphi^4$ theory. 

We first study an analogous relation for toy $D=0$ partition functions, which
are ordinary integrals. The relation in this case takes a form similar to
\eqref{e1} but without logarithms, and it can be rigorously proved because the
toy partition functions can be calculated exactly. We then discuss some simple
generalizations in $D=0$ in order to understand why the logarithms in~\eqref{e1}
appear when one considers partition functions with $D\geq1$, which are
functional integrals. For $D=1$, we partly check the relation by calculating the
imaginary contributions to the ground-state energy. Although on both sides of
\eqref{e2} the imaginary parts cancel trivially, before their cancellation one
can find agreement between separate pieces on the LHS and RHS. Since in the
path-integral formulation, there is no essential difference between the cases
$D>1$ and $D=1$ except for the well-understood regularisation and
renormalization, we conjecture that the relation~\eqref{e1} holds also for
$D>1$. 

\section{Case $D=0$}
\subsection{Full results}
\begin{figure}[ht]
\begin{center}
\includegraphics[scale=0.5]{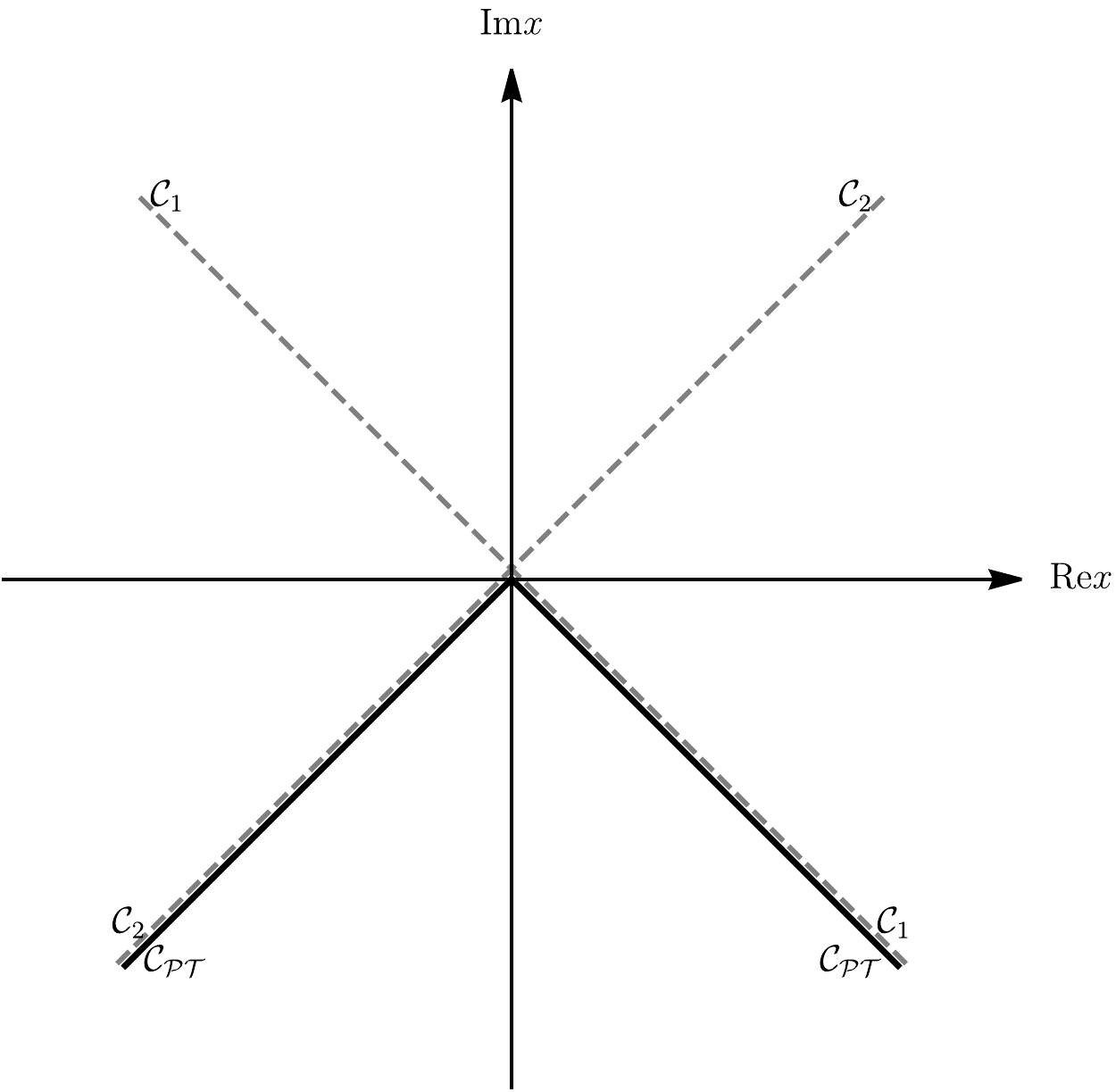}
\caption{An example of the contour $\C_{\P\T}$ is shown as the solid line. Also
shown are two other contours (dashed lines) $\C_1:\e^{-\i\pi/4 }(-\infty,+\infty
)$, and $\C_2:\e^{\i \pi/4}(-\infty,+\infty)$.}
\label{f1}
\end{center}
\end{figure}

The toy $D=0$ model has been studied earlier in~\cite{R04,R011} and is a good
platform to illustrate the idea. We first consider the Hermitian partition
function
\begin{align}
\label{e3}
\widetilde{Z}_{\rm Herm}(\lambda)=\int_{-\infty}^\infty
{\rm d}x\,\exp\left(-\half{m^2 x^2}-\fourth\lambda x^4\right),
\end{align}
where $\lambda>0$. When $\lambda=-g<0$, the integral above is divergent, but we can consider the \cPT-symmetric non-Hermitian theory,
\begin{align}
\label{e4}
\widetilde{Z}^{\P\T}(g)=\lim_{\epsilon\rightarrow 2}\int_{\C_{\P\T}}\d x\,
\exp\left(-\half m^2x^2-\fourth g x^2 (\i x)^\epsilon \right),
\end{align}
where for $\epsilon=2$, $\C_{\P\T}$ is a continuous contour that terminates in the \cPT-symmetric
Stokes wedges $-\frac{3}{8}\pi<\arg x<-\frac{1}{8}\pi$ and $-\frac{7}{8}\pi<
\arg x<-\frac{5}{8}\pi$. The contour $\C_{\P\T}$ is not unique; deforming a
contour joining the \cPT-symmetric Stokes wedges leaves the value of the
integral unchanged. An example of $\C_{\P\T}$ is shown as the solid line in
Fig.~\ref{f1}. Below, we show that the above toy partition functions satisfy
\begin{align}
\label{e5}
&\widetilde{Z}^{\P\T}(g)=\half\widetilde{Z}_{\rm Herm}(-g+\i 0^+)
+\half\widetilde{Z}_{\rm Herm}(-g-\i 0^+).
\end{align} 

Using the contour $C_{\P\T}$ shown in Fig.~\ref{f1} for \eqref{e4} we get
\begin{align}
\label{e6}
&\widetilde{Z}^{\P\T}(g)=\e^{\i\pi/4}\int_{-\infty}^0\d x\,\exp\left(-\half\i
m^2 x^2-\fourth gx^4\right)\notag\\
&\qquad\qquad\,+\e^{-\i\pi/4}\int^{\infty}_0\d x\,\exp\left(\half\i m^2 x^2-
\fourth gx^4\right)\notag\\
&=\frac{m\pi}{2\sqrt{2g}}\exp\left(-\frac{m^4}{8g}\right) \left[I_{-\frac{1}{4}}\left(\frac{m^4}{8 g}\right)+I_{\frac{1}{4}}\left(\frac{m^4}{8 g}\right)\right],
\end{align}
where $I_{\pm\frac{1}{4}}(z)$ is the modified Bessel function of the first kind. On
the other hand, doing the integral in~\eqref{e3} gives $\widetilde{Z}_{\rm Herm}(\lambda)=\frac{m}{\sqrt{2\lambda}}\exp\left(
\frac{m^4}{8\lambda}\right)K_{\frac{1}{4}}\left(\frac{m^4}{8\lambda}\right)$,
where $K_{\frac{1}{4}}(z)$ is the modified Bessel function of the second kind. $K_{\frac{1}{4}}(z)$ has a branch cut on the negative axis and for $z>0$
\begin{align}
\label{e7}
K_{\frac{1}{4}}(-z\pm\i 0^+)=&\frac{\pi}{2}\left(I_{-\frac{1}{4}}(z)-I_{\frac{1}{4}}(z)
\right)\notag\\
&\mp \frac{\i\pi}{2}\left(I_{-\frac{1}{4}}(z)+I_{\frac{1}{4}}(z)\right).
\end{align}
Substituting the above equation into $\half Z_{\rm Herm}(-g+\i 0^+)+\half Z_{\rm
Herm}(-g -\i 0^+)$, we obtain the result in \eqref{e6}, which proves the
relation~\eqref{e5}.

One may understand the relation~\eqref{e5} in another way. When the function
$\widetilde{Z}_{\rm Herm}(\lambda)$ is analytically continued away from the
positive real axis via $\lambda=g\exp(\i\theta)$, where $g>0$, it could still
have an integral representation as in \eqref{e3} but with the contour rotated
via $x\to x\exp(-\i\theta/4)$. Then, $\widetilde{Z}_{\rm Herm}(g\e^{\i(\pi-0^+)}
)$ could be represented by the same integral but with the contour given by
$\C_1$ in Fig.~\ref{f1}, while $\widetilde{Z}_{\rm Herm}(g\e^{-\i(\pi-0^+)})$
corresponds to the contour $\C_2$. In these cases, the integrand is the same as
that in~\eqref{e4} with $\epsilon=2$. Because of the symmetry $x\to -x$ in the
integrand, the contour $\C_{\P\T}$ can be effectively viewed as half $\C_1$ plus
half $\C_2$ and thus one arrives at the relation~\eqref{e5}.

\subsection{Imaginary parts and semi-classical estimates}\label{sec:II-B}
In the above example the partition functions can be calculated exactly. This is
not possible when $D\geq1$. Therefore, we search for an alternative approach
where precise results of the partition functions are not needed, but one still
can examine the relation between the Hermitian and the non-Hermitian
\cPT-symmetric Euclidean partition functions. From~\eqref{e7}, one sees that
$\widetilde{Z}_{\rm Herm}(-g\pm\i 0^+)$ contains an imaginary part:
\begin{align}
\label{e8}
& {\rm Im}\left[\widetilde{Z}_{\rm Herm}(-g\pm\i 0^+)\right]\notag\\
&=\pm\frac{m\pi}{2\sqrt{2g}}\exp\left(-\frac{m^4}{8g}\right)\left[I_{\frac{1}{4}}
\left(\frac{m^4}{8g}\right)-I_{-\frac{1}{4}}\left(\frac{m^4}{8g}\right)\right]
\notag\\
&=\mp \frac{\sqrt{\pi}}{m}\e^{-m^4/4g} {\rm\ for\ }g\to0.
\end{align}
Although these imaginary parts cancel trivially on the RHS of~\eqref{e5}, one
can still observe the same imaginary parts on the LHS in evaluating
$\widetilde{Z}^{\P\T}(g)$ semiclassically. 

For small $g$ we approximate the integral~\eqref{e4} using the method of
steepest descents~\cite{R012}. There are three stationary points: $x_{\rm L}=-m
/\sqrt{g}$, $x_0=0$, and $x_{\rm R}=m/\sqrt{g}$. Steepest paths satisfy the
constant-phase condition ${\rm Im}\left[m^2x^2/2-gx^4/4\right]={\rm const}$,
where the constant is determined by the phase at the stationary point. Writing
$x=x_{\rm Re}+\i x_{\rm Im}$, we have $m^2(x_{\rm Re} x_{\rm Im})=g(x_{\rm Re}
x_{\rm Im})(x^2_{\rm Re}- x^2_{\rm Im})$, which gives $x_{\rm Re}=0,\,x_{\rm Im}
=0,\,x^2_{\rm Re}-x^2_{\rm Im}=m^2/g$. At the stationary point $x_0$, the
solution with $x_{\rm Im}=0$ corresponds to the steepest-descent path, while at
$x_{\rm L}$ or $x_{\rm R}$, $x^2_{\rm Re}-x^2_{\rm Im}=m^2/g$ gives the
steepest-descent path. 

\begin{figure}[ht]
\begin{center}
\includegraphics[scale=0.5]{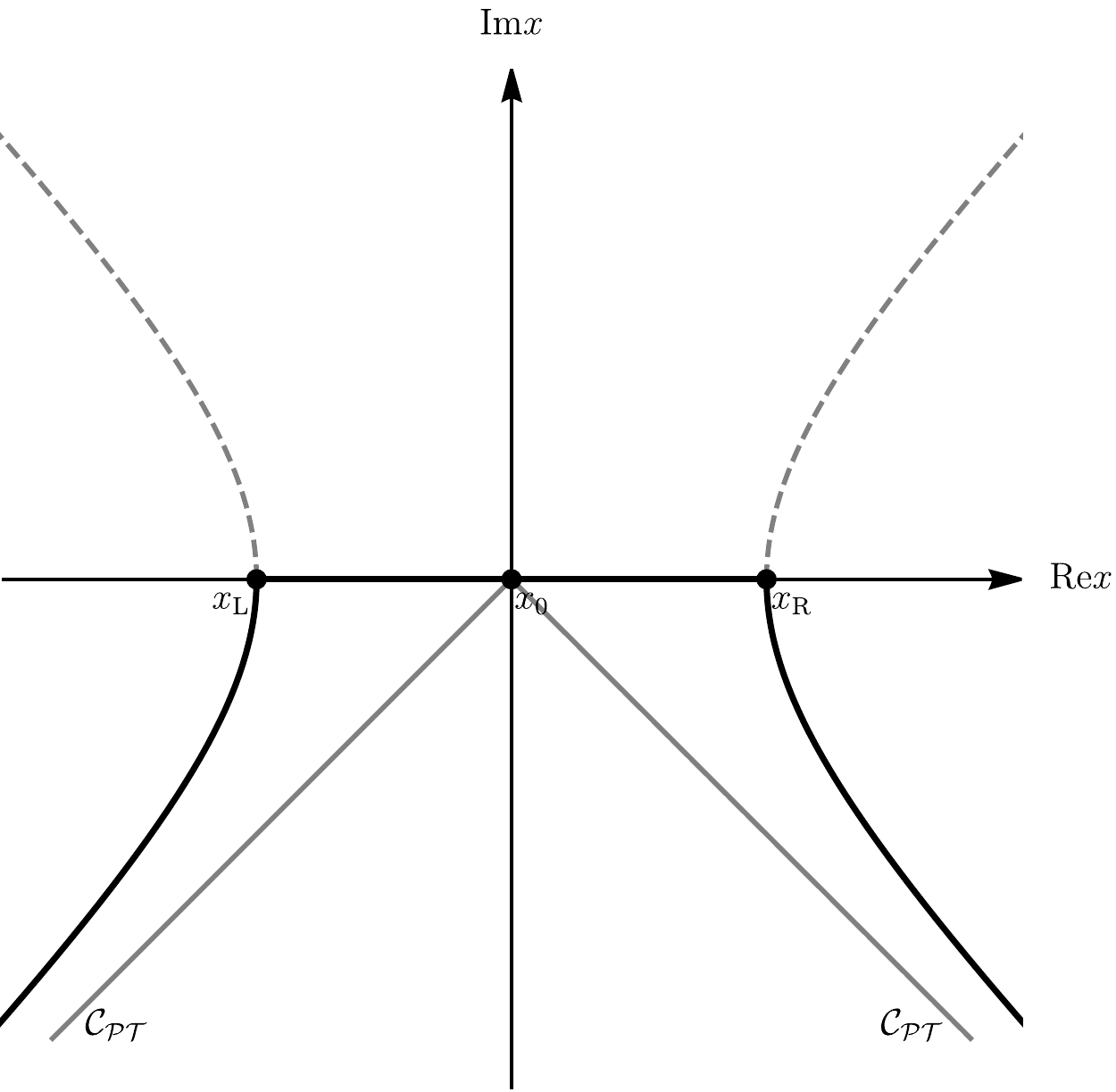}
\caption{The deformed contour $\C'_{\P\T}$ (heavy solid line) is composed of
steepest-descent paths through the stationary points $x_{\rm L}$ (left
half-hyperboloid), $x_0$ (the real line), and $x_{\rm R}$ (right
half-hyperboloid).}
\label{f2}
\end{center}
\end{figure}

Next, we deform the contour $\C_{\P\T}$ to the new one $\C'_{\P\T}$ shown in
Fig.~\ref{f2}. In the case of a path integral for $D\ge1$, all the
steepest-descent paths crossing a stationary point constitute a hypersurface
called a {\it Lefschetz thimble}~\cite{R013,R014,R015,R016}. For $D=0$,
Lefschetz thimbles reduce to the steepest-descent one-dimensional paths through the
stationary points, e.g., the hyperbolas in Fig.~\ref{f2}. Anticipating the generalisation
of our analysis to the case $D=1$, we call these paths in the present case {\it thimbles}, which are denoted by $\J_{\rm L}$,
$\J_0$, and $\J_{\rm R}$. We denote the half Lefschetz thimble of $\J_{\rm L}$
that goes to the lower (upper) half plane as $\J_{\rm L, -(+)}$. Similarly, we
denote the half Lefschetz thimble of $\J_{\rm R}$ going to the lower (upper)
half plane as $\J_{\rm R,-(+)}$. Thus, the deformed contour can be expressed as
$\C'_{\P\T}=\J_{\rm L,-}+\J_{0}+\J_{\rm R,-}$, which is left-right symmetric.
The leading contribution to $\rm{Im}[\widetilde{Z}^{\P\T} (g)]$ is easily obtained by
evaluating the integral on $\J_{\rm L,-}$ and $\J_{\rm R,-}$ up to the Gaussian
fluctuations:
\begin{equation}
{\rm Im}\left[\left.\widetilde{Z}^{\P\T}(g)\right|_{\J_{\rm L/R,-}}\right]
=\pm \frac{ \sqrt{\pi}}{2m}\exp(-m^4/(4g)).
\nonumber
\end{equation}
These imaginary parts differ by a sign and so they also cancel on the LHS
of~\eqref{e5}. Comparing the above equation with the last line in \eqref{e8}, we
see that for small $g$ the imaginary parts from $\J_{\rm L,-}$ and
$\J_{\rm R,-}$ are equal to those from $\half\widetilde{Z}_{\rm Herm}(-g-\i 0^+
)$ and $\half\widetilde{Z}_{\rm Herm}(-g+\i 0^+)$. The imaginary part from
$\J_{\rm L,-}$ is supposed to be that from the left part of the standard contour
$\C_{\P\T}$ which is also the left part of $\C_2$ (see Fig.~\ref{f1}). The
integral on the latter gives half of $Z_{\rm Herm}(-g-\i 0^+)$. This explains
the correspondence $\J_{\rm L,-}\leftrightarrow-g-\i 0^+;\, \J_{\rm R,-}
\leftrightarrow -g+\i 0^+$ in comparing the imaginary parts from the LHS and RHS
of~\eqref{e5}. 

\subsection{Simple generalizations}
The relation~\eqref{e5} is not general. To see why, consider a generalization of
the partition functions in~\eqref{e3} and~\eqref{e4}. One can consider $Z_{\rm
Herm}(\lambda)=[\widetilde{Z}_{\rm Herm}(\lambda)]^2$. This partition function
would be given by a double integral with the integrand being the product of that
in \eqref{e3} for each variable. The corresponding non-Hermitian partition
function would be given as $Z^{\P\T}(g)=[\widetilde{Z}^{\P\T}(g)]^2$. These
partition functions do {\it not} satisfy $Z^{\P\T}(g)=Z_{\rm Herm}(-g+\i 0^+)/2
+Z_{\rm Herm}(-g-\i 0^+)/2$ but rather $\sqrt{Z^{\P\T}(g)}=\sqrt{Z_{\rm Herm}(-g
+\i 0^+)}/2+\sqrt{Z_{\rm Herm}(-g-\i 0^+)}/2$. One can also consider $Z_{\rm
Herm}(\lambda)=\exp[\widetilde{Z}_{\rm Herm}(\lambda)]$ and $Z^{\P\T}(g)=
\exp[\widetilde{Z}^{\P\T}(g)]$, which satisfy~\eqref{e1} rather than~\eqref{e5}.

In these examples $\widetilde{Z}^{\P\T}$ and $\widetilde{Z}_{\rm Herm}$ are the
fundamental ingredients in the partition functions $Z^{\P\T}$ and $Z_{\rm
Herm}$. The relation between the Hermitian and non-Hermitian partition functions
should be expressed in terms of the fundamental ingredients. In the next
section we motivate that for realistic partition functions with $D\geq 1$, the fundamental ingredients are given by $\log Z$.

\section{Case $D=1$} 
For $D=1$, the Euclidean partition function is $Z=\int_\C \D x\,\exp\left(-S[\it{x}]
\right)$, where $S$ is the Euclidean action 
\begin{equation}
\nonumber
S\left[x \right]=\int\d\tau\left[ \frac{1}{2}\left(\frac{\d x(\tau)}{\d\tau}
\right)^2+V\left(x\right)\right].
\end{equation}
Usually, the partition function represents the Euclidean transition amplitude
between the ground state, $\langle 0|\e^{-\hat{H}T}|0\rangle=\int\d x_f\d x_i\,
\langle 0|x_f\rangle\langle x_f|\e^{-\hat{H}T}|x_i\rangle \langle x_i|0\rangle$,
where $\hat{H}$ is the Hamiltonian operator and $|x_{i/f}\rangle$ are position
eigenstates. The transition amplitude between position eigenstates $\langle x_f
|\e^{-\hat{H}T}|x_i\rangle$ can also be calculated from the Euclidean partition
function but with fixed boundary conditions $x(-T/2)=x_i$, $x(T/2)=x_f$ for the
functions to be integrated. 

The transition amplitude between position eigenstates is used in
practical calculations and one can project it onto the vacuum persistence
amplitude by taking $T\to\infty$. For the partition function $Z=\langle x_0|\exp
(-\hat{H}T)|x_0\rangle$ the imaginary part of the ground-state energy can be
calculated from this Euclidean partition function \cite{R017,R018,R019,R020}. To
see this let $\left\{|n\rangle,\ n\in\mathbb{N}\right\}$ denote a complete set of
energy eigenstates of $\hat{H}$ with energies $E_{n}$ increasing with $n$. We
then have $Z=\langle x_0|\exp(-\hat{H}T)|x_0\rangle=\sum_{n} \exp(-E_n T)|
\langle x_0|n\rangle|^2$. Taking the imaginary parts on the logarithms of both sides as $T\to
\infty$, we get
\begin{align}
\label{e9}
{\rm Im}\,E_{0}=-\lim_{T\to\infty}\frac{{\rm Im}\left[\log Z\right]}{T}.
\end{align}
Of course, if the theory has a stable ground state, then ${\rm Im}\,E_0=0$. For the potentials of interest here, one may choose $x_0=0$ as this point has the highest weight in the ground-state wave function. For the Hermitian theory
with $V(x)=\half m^2 x^2+\fourth\lambda x^4$ ($\lambda>0$), this defines $Z_{\rm
Herm}(\lambda)$ and the contour $\C$ is in real function space: $\{x(\tau): x(-
T/2)=x(T/2)=0\}$. For the non-Hermitian \cPT-symmetric theory with $V(x)=\half m^2x^2-\fourth gx^4$ ($g>0$), one must assign a contour $\C_{\P\T}$ properly to define $Z^{\P\T}(g)$. A possible way to define $\C_{\P\T}$ is to let $x=\theta(-s)\exp(\i\pi/4)s+\theta(s)\exp(-\i\pi/4)s$, where $s$ is real and $\theta(s)$ is the step function. Then $\C_{\P\T}$ is composed of all the real
functions $s(\tau)$ with $s(-T/2)=s(T/2)=0$.

Unlike the $D=0$ case, the \cPT-symmetric partition function defined above is
difficult to compute. However, from the insights obtained in the $D=0$ case,
we conjecture the relation~\eqref{e1} between $Z^{\P\T}(g)$ and $Z_{\rm Herm}
(\lambda)$ so that the former can be calculated from the latter. Below, we adopt
the semiclassical evaluation of $Z^{\P\T}(g)$ to motivate and partly check this
relation.

\subsection{Semiclassical evaluation of $Z^{\P\T}(g)$}
As in Sec.~\ref{sec:II-B}, to evaluate $Z^{\P\T}(g)$ semiclassically we deform
the contour $\C_{\P\T}$ and apply the method of steepest descents. We call the
real function space $\{x(\tau): x(-T/2)=x(T/2)=0\}$ the {\it real hyperplane},
denoted as $\mathbb{R}^\infty$. This is an infinite-dimensional space. To apply
the method of the steepest descents we need to complexify this space to
$\{z(\tau): z(-T/2)=z(T/2)=0\}$, where $z(\tau)$ are complex functions. We
denote the latter as $\mathbb{C}^{\infty}$. We then find the deformed contour
$\C'_{\P\T}$ that passes through all relevant stationary points. The deformed
contour $\C'_{\P\T}$ is a middle-dimensional hypersurface~\cite{footnote1} in the complexified space $\mathbb{C}^{\infty}$.

First, we identify the stationary points and we assume that all relevant
stationary points are real as in the simple $D=0$ case. In the present case, we
have infinitely many real stationary points, but among all stationary points
only three are {\it fundamental} (in the so-called dilute-instanton-gas approximation) and are in one-to-one correspondence with the
$D=0$ stationary points. All others are composed of these fundamental stationary
points. First, we have the
trivial stationary point $x_0(\tau)=0$. We also have two fundamental instantons
(also called {\it bounces} in the context of false-vacuum
decay~\cite{R017,R018}), $x_{\rm L}(\tau)$ and $x_{\rm R}(\tau)$, whose explicit
forms are determined by solving the equation of motion
\begin{equation}
\nonumber
-\frac{\d^2 x}{\d\tau^2}+m^2 x-gx^3=0
\end{equation}
with boundary conditions $x(\tau)|_{\tau\to\pm\infty}=0$ and
$\d x/\d\tau|_{\tau=\tau_0}=0$. The solutions are
\begin{eqnarray}
x_{\rm L}(\tau;\tau_0)&=&-\frac{m}{\sqrt{g}}\sqrt{2\left(1-{\rm tanh}^2[m(\tau
-\tau_0)]\right)},\nonumber\\
x_{\rm R}(\tau;\tau_0)&=&\frac{m}{\sqrt{g}}\sqrt{2\left(1-{\rm tanh}^2[m(\tau
-\tau_0)]\right)}.\nonumber
\end{eqnarray}
Note that the factors $\mp m/\sqrt{g}$
are simply the stationary points $x_{\rm L}$, $x_{\rm R}$ in the simple $D=0$
toy partition functions.

The free parameter $\tau_0$ is a collective coordinate of the bounce
characterizing its ``position''. In a rigorous sense, the parameter $\tau_0$
means that there are infinitely many fundamental bounces for each type ``L/R''
that are degenerate in the Euclidean action. For any solution with chosen
$\tau_0$, the time invariance is spontaneously broken and thus there is a zero
mode in the fluctuations about a chosen bounce solution. The integral over the fluctuations in the zero-mode direction can be traded for an
integral over the collective coordinate; this amounts to adding up all the
degenerate fundamental bounces of the same type.

Now each bounce solution can be viewed as a point on the real hyperplane as
long as the degeneracy characterized by $\tau_0$ is taken into account properly.
Like the zero-dimensional case associated with $x_0(\tau)$, $x_{\rm L}(\tau;
\tau_0)$, and $x_{\rm R}(\tau;\tau_0)$, we have three relevant Lefschetz
thimbles $\J_{\rm L}$, $\J_0$, and $\J_{\rm R}$ that are composed of the
steepest-descent paths passing through them. In general, the steepest-descent
paths passing through a stationary point $z_a(\tau)$ can be obtained by solving
the gradient {\it flow} equation~\cite{R013}
\begin{align}
\nonumber
\frac{\partial z(\tau;u)}{\partial u}=\overline{\frac{\delta {S[z(\tau;u)]}}
{\delta {z(\tau;u)}}}
\end{align}
and its complex conjugate. Here $u\in \mathbb{R}$ and the boundary condition is
$z(\tau; u=-\infty) = z_a(\tau)$. Denoting $h[z]=-{\rm Re}\,S[z]$, one can
check that $\partial h/\partial u\leq 0$. The real part of $-S[z]$ decreases as
we move away from the stationary point along the path given by $z(\tau;u)$
(see Refs.~\cite{R013,Ai:2019fri}).

The present situation is much simpler. To identify the steepest-descent paths
passing through the fundamental stationary points we consider the eigenvalue
equation on the real hyperplane
\begin{align}
\nonumber
\left[-\partial^2_\tau+V''(x_a(\tau))\right]f^a_i(\tau)=\lambda^a_i f^a_i(\tau),
\end{align}
where $a=0,{\rm L}, {\rm R}$. The fluctuation operators $-\partial_\tau^2+V''(
x_a)$ are generalizations of Hessian matrices. In the tangential space near
$x_a(\tau)$ the eigenfunctions $\{f_i^a\}$ provide a basis. For the cases
$\lambda_i^a>0$ (positive mode), $\lambda_i^a=0$ (zero mode), and $\lambda_i^a<
0$ (negative mode), $\pm f_i^a$ generates two downward, flat, and upward paths
with respect to the ``height function'' $h[x]=-{\rm Re}S[x]$. (Here the two
paths generated by $\pm f_i^a$ join at the stationary point $x_a(\tau)$ and form
a continuous curve.) Therefore, non-negative modes generate steepest-descent
paths that still lie on the real hyperplane of the configuration space. For a
negative mode one must look into the subcomplex plane whose real axis is
generated by that negative mode. Associated with the negative mode, the
steepest-descent paths leave the stationary point and go to the upper and lower
imaginary directions on that subcomplex plane, in analogy with the hyperbolas in
Fig.~\ref{f2}.

It is well known in the context of false vacuum decay~\cite{R018} that
there is only one negative mode for the fluctuations about each bounce solution
while there is no negative mode about the trivial solution. Therefore, all
steepest-descent flows passing through $x_0(\tau)$ lie on the real hyperplane. For
$\J_{\rm L}$ ($\J_{\rm R}$), there are two steepest-descent paths that do not
lie on the real hyperplane. Similarly, we only pick one of them, defining
$\J_{\rm L/R,-}$ and $\J_{\rm L/R,+}$. 

Aside from the two fundamental bounces, there are {\it multiple-bounce}
solutions that form stationary points, whose contribution to the path integral
in the dilute-instanton approximation is a combination of $n_1$ left-type
bounces and $n_2$ right-type bounces, which are separated by intervals much
larger than the duration of each single bounce~\cite{R018,Ai:2020vhx}. We label
these multibounces by $(n_1,n_2)$. Denote the partition function evaluated on
$\J_{\rm L/R,-}$ near the single-bounce $x_{\rm L/R}$ including the collective
coordinate integrated over as $Z^{\P\T}_{{\rm L/R},-}$. Then the partition function
evaluated on the thimble $\J_{(n_1,n_2),-}$ passing through the multibounce
$(n_1,n_2)$ has the form
\begin{align}
Z_{(n_1,n_2),-}^{\P\T}=\frac{Z_0^{\P\T}}{n_1!n_2!}\left(\frac{Z_{\rm
L,-}^{\P\T}}{Z_0^{\P\T}}\right)^{n_1}\left(\frac{Z_{\rm R,-}^{\P\T}}
{Z_0^{\P\T}}\right)^{n_2},
\nonumber
\end{align}
where $Z_0^{\P\T}$ is the partition function evaluated on $\J_0$ near $x_0$ and
its appearance is due to the contribution from the trivial configurations
between any two neighboring bounces. The factor $n_1!$ or $n_2!$ is due to the
symmetry when exchanging the positions of the bounces of the same type. Then for an $n$-bounce, we have
\begin{align}
Z^{\P\T}_{n,-}=\sum_{n_1+n_2=n} Z^{\P\T}_{(n_1,n_2),-}. \nonumber
\end{align}
The full partition function can be expanded as
\begin{align}
Z^{\P\T}&=Z_0^{\P\T}+\sum_n Z_{n,-}^{\P\T}\notag\\
&=\exp\left(\frac{Z^{\P\T}_{\rm L,-}}{Z_0^{\P\T}}
+\frac{Z^{\P\T}_{\rm R,-}}{Z_0^{\P\T}}+\log Z_0^{\P\T}\right).
\label{e10}
\end{align}
In the full $Z^{\P\T}$, all the fundamental stationary points are completely
entangled with each other because of the multibounce configurations. In $\log
Z^{\P\T}$, they ``decouple'' from each other and have one-to-one correspondence
to the three stationary points in the simple $D=0$ case (see Fig.~\ref{f2}).
Similarly, the Hermitian partition function $Z_{\rm Herm}$ can also be put in an exponential form with $\log Z_{\rm Herm}$ playing a fundamental role; the latter
is given by connected Feynman diagrams. This, together with insights obtained
from the $D=0$ case, motivates us to conjecture the relation~\eqref{e1} between
$\log Z^{\P\T}$ and $\log Z_{\rm Herm}$. This relation indicates that the energy
of the non-Hermitian $-gx^4$ theory can be calculated from the Hermitian
$\lambda x^4$ theory via \eqref{e2}. 

Below, we partly check \eqref{e2}, and also relation~\eqref{e1}, by comparing
the imaginary parts on the LHS and RHS for the ground-state energy.
Substituting~\eqref{e10} into~\eqref{e9} gives
\begin{align}
\label{e11}
{\rm Im} E^{\P\T}_0=-\lim_{T\to\infty}\frac{1}{T}{\rm Im}\,\left(\frac
{Z^{\P\T}_{\rm L,-}}{Z_0^{\P\T}}+\frac{Z^{\P\T}_{\rm R,-}}{Z_0^{\P\T}}\right),
\end{align}
where we have used ${\rm Im}\log Z_0^{\P\T}=0$ because the integral $Z^{\P\T}_0$
is performed on the real hyperplane near the trivial stationary point.

\subsection{Imaginary parts from nonperturbative stationary points}
We now evaluate $(Z^{\P\T}_{\rm L/R,-}/Z_0^{\P\T})$. To integrate over the Gaussian fluctuations on a Lefschetz thimble about a saddle point, one usually needs to solve the flow equations that determine the tangential space about the stationary point on the thimble~\cite{R013}. However, as mentioned above,
our case is much simpler and the well-known formula for false-vacuum decay rates
applies~\cite{Ai:2019fri}. The integral over fluctuations can be expressed
in terms of functional determinants of the Euclidean fluctuation operators:
\begin{align}
&\frac{Z_{\rm L/R,-}^{\P\T}}{Z_0^{\P\T}}=\pm\left(\frac{\i}{2}\right)T
\left(\frac{S_{\rm E}[x_{\rm L/R}]}{2\pi}\right)^{1/2}\notag\\
&\quad\;\times\left|\frac{{\det}'\left[-\partial_\tau^2+V''(x_{\rm L/R}(\tau))
\right]}{{\det}\left[-\partial_\tau^2+V''(x_0)\right]}\right|^{-1/2}
\e^{-S_{\rm E}[x_{\rm L/R}]}.
\label{e12}
\end{align}
Here, the prime on $\det$ indicates that the zero mode is excluded when evaluating the
functional determinant. The integral over the fluctuations in the zero-mode
direction can be traded for an integral over the collective coordinate
\cite{R023}, giving the factor $T\sqrt{S_{\rm E}[x_{\rm L,R}]/{2\pi}}$ in the
above equation, where ``+'' corresponds to the left-type bounce and ``-'' to
the right-type bounce.

The functional determinants can be calculated using the powerful Gel'fand-Yaglom
theorem~\cite{R024}. The results with the zero modes removed are given in
Ref.~\cite{R019} for the kink-type solutions and in Ref.~\cite{Ai:2019fri} for
the bounce-type solutions. For our case, we have
\begin{align}
\frac{{\det}'\left[-\partial_\tau^2+V''(x_{\rm L/R}(\tau))\right]}{{\det}
\left[-\partial_\tau^2+V''(x_0)\right]}=-\frac{1}{12 m^2}.
\notag
\end{align}
Substituting the above result into \eqref{e12}, we obtain 
\begin{equation}
\label{e13}
-\lim_{T\to\infty}\frac{1}{T}{\rm Im}
\left(
\frac{Z^{\P\T}_{\rm L/R,-}}
{Z_0^{\P\T}}
\right)
=\mp m^2\sqrt{\frac{2m}{\pi g}}\e^{-4 m^3/(3g)}.
\end{equation}
Below, we observe the same imaginary parts from the RHS of \eqref{e2}.

\subsection{Hermitian perturbation theory}
The energy for the Hermitian theory can be expressed as a
Rayleigh-Schr\"{o}dinger perturbation series~\cite{R025}. In particular, for the
ground-state energy, we have
\begin{align}
\label{e14}
E_{0,\rm Herm}(\lambda)=\frac{m}{2}+\frac{m}{2}\sum_{n=1}^\infty A_n
\left(\frac{\lambda}{4m^3}\right)^n\,,
\end{align}
where $A_n$ have the asymptotic behaviors $A_n\sim \sqrt{6}(-1)^{n+1}3^n\Gamma
(n+\half)/\pi^{3/2}$. Taking the Borel sum of the above series and analytically
continuing $\lambda$ to $-g\pm\i 0^+$ ($g>0$) would give rise to imaginary parts
in $E_{0,\rm Herm}(-g\pm \i 0^+)$. These imaginary parts are only sensitive to
the large-order behavior of $A_n$. Therefore, we consider a series, denoted by $F_0$, having the
same form as \eqref{e14} but with $A_n=\sqrt{6}(-1)^{n+1}3^n\Gamma(n+\half)/
\pi^{3/2}$ for all $n\geq1$. One then has ${\rm Im}
F_0(\lambda)={\rm Im}E_{0,\rm Herm}(\lambda)$. $F_0(\lambda)$ reads
\begin{align}
F_{0}(\lambda)=\frac{m}{2}-\frac{m\sqrt{6}}{\pi^{3/2}}\int_0^\infty\d t\,
t^{-\frac{1}{2}}\e^{-t}\left[\frac{1}{1+\frac{3\lambda t}{4m^3}}-1\right],
\notag
\end{align}
which gives
\begin{align}
{\rm Im}F_0(-g\pm \i 0^+)=\pm 2m^2\sqrt{2m/(\pi g)}\e^{-4m^3/3g}\,.
\notag
\end{align}
Finally, we get
\begin{align}
\half{\rm Im}\,E_{\rm 0,Herm}(-g\pm \i 0^+)=\pm
m^2\sqrt{2m/(\pi g)}\e^{-4m^3/3g},
\notag
\end{align}
which are precisely the same as those from \eqref{e13} and therefore from
${\rm Im}E_0^{\P\T}(g)$ per \eqref{e11}. Thus, there is indeed a correspondence
between the imaginary parts on the LHS and RHS in \eqref{e2} applied to the
ground-state energy. Note that we again have the correspondence $\J_{\rm L,-}
\leftrightarrow-g-\i 0^+;\, \J_{\rm R,-}\leftrightarrow -g+\i 0^+$ in comparing
the imaginary parts.

To generalize the above analysis to $D>1$, one can simply replace $x(\tau)\to
\varphi(x)$ and use the corresponding higher-dimensional action. Again, there
are three fundamental bounce solutions $\varphi_0(x)$, $\varphi_{\rm L}(x)$, and
$\varphi_{\rm R}(x)$. The previous analysis is still valid, so we conjecture
that the relation~\eqref{e1} also holds for $D>1$ \cite{R026}.

\section{Conclusions}
In search of physics beyond the Standard Model of particle physics, the use of
non-Hermitian Hamiltonians has only recently been used in model building
\cite{R01}. In this paper we have proposed a new approach to study non-Hermitian
\cPT-symmetric theories, in which one searches for relations between quantities
in the non-Hermitian and the corresponding Hermitian theories. We have focused
on the partition functions for the $-g\varphi^4$ theory, but there is no reason
in principle, why similar relations for other theories (for example, for
$\epsilon\neq 2$) and for other quantities cannot be constructed. This approach opens a new avenue to explore non-Hermitian \cPT-symmetric theories. 

The path-integral formulation we have adopted to build the central relation
\eqref{e1} is very general, and the relation may hold for spacetime dimension $D\geq1$.
This relation immediately implies that the energy spectrum for the
\cPT-symmetric $-g\varphi^4$ theory is real. Of course, \eqref{e1} remains a
conjecture that has only been partly checked by comparing the imaginary parts
on the LHS and RHS of \eqref{e2} for the ground-state energy. Our analysis is also related to work on resurgence and the analysis of large-order behavior in perturbation theory~\cite{R25}. Given the challenge, it is important to use all complementary approaches to understand the predictions
concerning \cPT-symmetric field theory in $D=4$ spacetime.

\section*{Acknowledgments}
WYA, CMB, and SS are supported by the UK Engineering and Physical Sciences
Research Council under research grant EP/V002821/1. CMB is also supported by
grants from the Simons and the Alexander von Humboldt Foundations.


\begin{thebibliography}{99}
\bibitem{footnote1} 
The integration $n$-form in real space, when analytically continued to complex space, leads to a $2n$-form; to avoid this we need to restrict the $2n$-form to a $n$ (middle) dimensional differential form.
\end{thebibliography}
\end{document}